\documentstyle[aps,preprint]{revtex}
\begin{document}
\draft
\title{The effect of electro-magnetic formfactors on dilepton production
off pp-collisions}
\draft
\author{F. de Jong and U. Mosel}
\address{Institut f\"ur Theoretische Physik, Universit\"at Giessen,
35392 Giessen, Germany}
\date{\today}
\maketitle
\begin{abstract}
We investigate the effect of the electric formfactor on dilepton production
off pp-collisions. In the model we use, the formfactor is
separated in a 
pion-cloud contribution and a Vector Meson Dominance part. We find 
that the dilepton spectrum is hardly affected by the pion-cloud 
part, but significantly by the VMD contribution. 
We point out that data with sufficient resolution would provide a crucial
test of the validity of VMD for the nucleon formfactor in the timelike 
region. 
\end{abstract}

\pacs{}
The electro-magnetic formfactor of the nucleon has been a subject of 
study for many years.
The only direct experimental measurements of this formfactor are in the 
on-shell (both the incoming and outgoing nucleon are on-shell) spacelike
region, and in the timelike region above the $N\bar{N}$ annihilation 
threshold.
While the spacelike part -- at not too high photon momenta -- is
described by the so-called dipole fit, the timelike region below 
the $N \bar{N}$ threshold can be accessed only by analytically continuing
pole-fits based on the assumption of Vector Meson Dominance
\cite{Hoehler,Mergell}.
This  procedure leads to the prediction of very strong resonance-like 
structures at the vector meson masses which, however, are all in the 
physically forbidden region for on-shell nucleons.
In this letter we explore the possibility to experimentally check this
behavior by determining the half-offshell formfactor 
through studies dilepton production off nucleon-nucleon 
collisions. 
This process provides a means for studying several off-shell effects. 
Apart from the off-shell electro-magnetic formfactors also 
the off-shell behavior of the nucleon-nucleon interaction enters in
the problem. 
We studied this aspect in a previous publication \cite{FdJ_dilept}.
Apart from the interest in the fundamental process, dilepton production off
nucleon-nucleon collisions are also of interest in heavy-ion collisions
where they provide an important background in the dilepton spectrum. 

Various efforts have been made in the theoretical description of the
electro-magnetic formfactor, either on a 
semi-phenomenological level \cite{Iachello,Brown} or in a
microscopical approach \cite{Naus,Tiemeijer,Doenges}.
The latter models generally describe the electromagnetic vertex in terms of 
the (virtual) photon coupling to a bare nucleon surrounded by a 
mesonic cloud.
Another important ingredient of the formfactors is vector meson dominance
(VMD). 
The idea was first proposed by Sakurai \cite{Sakurai}, it presumes the 
photon couples to the hadron by first converting into a vector meson. 
The concept turned out to be extremely successful in the description of the
on-shell formfactor of the pion.
In the present work we will use the formfactors as calculated in 
Ref. \cite{Doenges}, these give a very good fit to the on-shell 
formfactors.

Data for dilepton production off nucleon-nucleon collisions are rather sparse.
For the elementary reaction we are only aware of the 
DLS data at a beam-energy of 4.9 MeV.  
However, at this energy there other processes, e.g. the Dalitz decays, are
more important then the 'pure' bremsstrahlung process and therefore 
these data are not very well suited for our aims.  
In the kinematical region we are interested in, 
the closest to an elementary reaction are the proton-Beryllium 
data at a laboratory energy of 2.1 GeV from Ref. \cite{Naudet}. 
This cross-section contains contributions from proton-proton and 
proton-neutron collisions. 
The latter are a factor of 2.5 larger (calculated in a model
without formfactors \cite{Schaefer}) at this laboratory energy and 
dominate the cross-section. 
The reason for this is a destructive interference between the various 
$pp$ diagrams, an effect not present in the $np$ diagrams.
Using two simplified models of the formfactors, the authors of 
Ref. \cite{Schaefer} find significant effects of the inclusion of
simple, VMD-like, 
formfactors on the cross-section of this specific process,
significantly overpredicting the data. 
Since the on-shell electromagnetic formfactor of the pion is well known
(also in the spacelike region), its inclusion is 
rather unambiguous apart from possible off-shell effects.
For the nucleon formfactors the situation is much less clear;
the results of Ref. \cite{Doenges} suggest that one may have to go beyond
a simple minded VMD approach to avoid the overprediction of the data. 
In order to concentrate on this point and to explore the effects of an off-shell
dependence of the formfactors we study in this paper dilepton
production off proton-proton collisions. 
This provides the cleanest probe of effects of nucleon electro-magnetic 
formfactors.

For the calculation of the differential cross-section of dilepton production
we use the model of Schaefer et al. \cite{Schaefer}.
The diagrams included in this model are shown in Fig. \ref{diagrams},
the intermediate state can either be a nucleon or a $\Delta$-isobar.
As e.g. shown in Refs. \cite{Schaefer,FdJ_dilept}, the dominant contributions
stem, especially at lower laboratory energies and smaller invariant 
masses of the dilepton pair, from the diagrams where the intermediate state 
is a $\Delta$-isobar.
In the model of Schaefer et al. the NN interaction, represented by the 
T-matrix in Fig. \ref{diagrams}, is parametrized in terms of an 
effective One-Boson-Exchange interaction. 
This parametrization is then extrapolated to find the off-shell matrix 
elements needed in the calculation of the dilepton cross-section. 
The fitted interaction reproduces the on-shell data well from a 
laboratory energy of 800 MeV up to 3.2 GeV.
This makes it well suited for calculations where one needs the NN T-matrix
at high energies, since at these energies  
the more elaborate models that calculate a full T-matrix in a potential
model have problems in describing the on-shell data.
Another advantage of the OBE-parametrization is that one obtains a current
that is gauge-invariant and one does not have the problem of devising 
a (non-unique) scheme to repair gauge invariance as we had to do in 
Ref. \cite{FdJ_dilept} where we used a realistic T-matrix based on a
potential model.
On the other hand we have the disadvantage that one has only limited control
over the off-shell behavior of the fitted NN interaction. 
For example, one finds a considerable difference between the obtained
spectra when taking either a pseudo-scalar of a pseudo-vector coupling
for the pion in the OBE interaction. These two couplings are only 
equivalent on-shell, where the interaction is fitted to the NN data, 
and the fitted coupling constants are thus the same for both choices.
Using the effective $T$-matrix one also misses the rescattering diagrams.
As we showed in Ref. \cite{FdJ_dilept} the difference between the present
model and a more elaborate one where we used a realistic T-matrix 
and did include the rescattering diagrams are rather marginal. 

Another difference between these two approaches is the treatment of 
negative energy states. 
In the potential model approach they are explicitly excluded. 
To incorporate them in this approach one needs a $T$-matrix which
dynamically includes negative energy states. 
These type of models are very involved and presently they
are only available for energies below pion threshold \cite{Fleischer,Gross}. 
In Ref. \cite{FdJ_ppg_neg} we presented a calculation for real
photon bremsstrahlung using the model of Ref. \cite{Gross}. 
We found that the various negative-energy state contributions
cancel almost totally, leaving the cross-section unaffected. 
Using the same model for dilepton production we 
find a similar cancellation and a negligible effect on the
cross-section. 
In the OBE based approach negative energy states are present;
they are essential for exact gauge invariance. 
However, their contribution to the cross-section is small. 

As can be inferred from Fig. \ref{diagrams} only the half-off-shell
formfactors enter in the calculation since either the incoming or outgoing
nucleon of the vertex that emits the virtual photon on-shell.
In this work we will use the ones calculated by Doenges et al. in which the
nucleon is described as a point source surrounded by a pionic 
cloud \cite{Doenges}. 
This physical picture is implemented by calculating all one-loop diagrams
with nucleon, $\Delta$ and pionic degrees freedom.

We first will shortly review the formalism used by Doenges. The most
general form of the nucleon-nucleon-photon vertex is \cite{Bincer}
(ignoring the magnetic part, we set this equal to the on-shell value of 
magnetic moment in the rest of the calculation):
\begin{equation}
\Gamma^\mu = \sum_{r,r'=0,1} (\not \! p')^{r'}
(A_1^{r' r} \gamma^\mu + A_3^{r' r} q^\mu) (\not \! p)^r.
\end{equation}
The coefficients $A_1$ and $A_3$ are functions of the three kinematical
variables in the problem: the incoming momentum $p^2$, the outgoing momentum
${p'}^2$ and the photon momentum $q^2$. Note that not all possible 
combinations of $p^2, {p'}^2, q^2$ are kinematically allowed. 
One can use the Ward-Takahashi identity to uniquely express the $A_3^{r'r}$ 
in terms of the various $A_1^{r'r}$, thereby ensuring gauge-invariance for the
calculated amplitude.
This will be done implicitly and in the following we concentrate on the 
various $A_1^{r'r}$ coefficients.

A different formulation is obtained by projecting on positive/negative 
energy content. 
The advantage of this is the relative ease with which one can express the 
reducible form-factors in terms of the irreducible ones. The latter are
the natural result of theoretical calculations like the one of Doenges et al. 
In this basis we have for the vertex:
\begin{eqnarray}
\Gamma^\mu &=& \sum_{s,s' = \pm} \Lambda^{s'}(W')
F_1^{s's} \Lambda^{s}(W) \nonumber \\
&\mbox{with }& \Lambda^\pm(W) = 
\frac{W \pm \not \! p}{2 W}, W^2 = p^2.
\end{eqnarray}
There is a straightforward one-one relation between the four coefficients
$A_1^{r'r}$ and the four form-factors $F_1^{\pm \pm}$, the most straightforward
being that when we have only the bare coupling, ($\Gamma^\mu = \gamma^\mu$),
we find $F_1^{\pm \pm} = 1$.
In our model only half off-shell formfactors enter: either the incoming
or the outgoing nucleon is on-shell ($W = m_N$).
This results in two of the four formfactors dropping out of the 
problem since $F_1^{\pm -}(W',m_N,q^2)u(p) = 0$. 
Together with the relation 
$F_1^{\pm +}(W',m_N,q^2) = F_1^{+ \pm}(m_N, W',q^2)$ we are left with 
only two formfactors. 
We then can work out the half off-shell vertex functions further:
\begin{equation}
\Gamma^\mu u(p) = \left[
\gamma^\mu \frac{F_1^{++} + F_1^{-+}}{2} + 
\not \! p' \gamma^\mu \frac{F_1^{++} - F_1^{-+}}{2W'}
\right] u(p).
\label{ff_half}
\end{equation}
At first glance it might appear that we have a problem when ${p'}^2 \leq 0$
(a kinematical region reached in our calculations) leading to 
an imaginary value of $W'$. 
However it turns out that 
$F^{++}({W'}^2 \leq 0) = {F^{-+}}^*({W'}^2 \leq 0)$, giving the special
case
\begin{equation}
\Gamma^\mu({W'}^2 \leq 0) u(p) = 
\gamma^\mu \frac{Re[F_1^{++}({W'}^2 \leq 0)]}{2} + 
\not \! p' \gamma^\mu \frac{Im[F_1^{++}({W'}^2 \leq 0)]}{2 |W'|}.
\end{equation}
For ${W'}^2 \rightarrow 0$ the limit exists due to the fact that
$Im[F_1^{++}({W'}^2 = 0)] = 0$. Finally we note that for real photons
we have $F^{\pm +}(W', q^2 = 0) = 1$, which implies that with real photons
one cannot measure half off-shell formfactors.
It also means that, generally speaking, we can expect to find larger effects 
of the formfactors for larger $q^2$.
This again implies one has to strive for large laboratory energies with which 
larger values of $q^2$ can be reached. 
On the other hand, at too high energies other processes become important
which again dilutes possible effects of formfactors.

As mentioned above, vector meson dominance is an important ingredient
in the formfactor.
For the nucleons, however, it can only partly describe the formfactor:
the on-shell formfactor in the spacelike region ($q^2 < 0$) has a 
dipole shape whereas the pure VMD hypothesis leads to a monopole 
formfactor. 
In the model of Doenges the VMD formfactor is included by adding the
appropriate photon-vector-meson coupling terms to the Lagrangian.
The resulting formfactor is added to the contributions from the pion-cloud 
giving a total formfactor:
\begin{equation}
F^{\pm \pm}_1(W', q^2) = F^{\pm \pm}_{1,\mbox{cloud}}(W', q^2) + 
\frac{1}{2} \frac{q^2}{m_{\omega}^2 - i m_\omega \Gamma_\omega(q^2) - q^2} +
\frac{1}{2} \frac{q^2}{m_{\rho}^2 - i m_\rho \Gamma_\rho(q^2) - q^2},
\label{full_ff}
\end{equation}
where the $\Gamma$'s are the momentum dependent widths of the mesons.
Note that the complex phase between the $\rho$- and $\omega$-pole terms
in Eq. \ref{full_ff} has been neglected by Doenges.
This choice gives a good fit to the pion formfactor, however, for a more 
precise fit one needs to include these phases \cite{om_phase}.
Recall again that the $F^{\pm \pm}_3$ formfactors are determined via the 
Ward-Takahashi identity.
Furthermore, in \cite{Doenges} the formfactors were extracted from 
physical transition amplitudes. 
Therefore the observable quantities, like cross-section, are representation
independent.

The resulting total nucleon formfactors provide an excellent description of 
the experimental values for the electric formfactor in the spacelike 
region \cite{Doenges}. 
Moreover, in the spacelike region the $F^{\pm \pm}_{1,\mbox{cloud}}$ are
significantly different from $1$ and not equal to each other, which leads
to a vertex notably different from the bare one. 
In the timelike region the deviations from $1$ are smaller, 
so that the vector meson pole terms dominate the spectrum there, independent
of $W'$.
Although all possible thresholds of pion production are included, these
do not clearly show up in the calculated reducible formfactors. 
This is mainly due to the Ward-Takahashi identity, which leads to 
cancellations: changes in the irreducible vertex are countered by
similar changes due to self-energy corrections in the reducible 
vertex \cite{Doenges}. 
We note here also that the formfactors are calculated for both positive
and negative $q^2$ for the same analytical model. The results therefore
respect the dispersion relations between the real and imaginary part.
The description of the magnetic formfactor is less satisfactory, only 
after rescaling the calculated formfactor to reproduce the value
at $q^2 = 0$ ($\kappa_p = 1.79$ one reproduces the data.
We therefore did not include the magnetic formfactors and set the magnetic
part of the vertex equal to its value at $q^2 = 0$.

On the $N \Delta \gamma$ vertices we only include a VMD-type factor.
Due to the isospin structure only isospin-1 mesons can couple and  
we only have a contribution of the $\rho$-meson.
So, when assuming VMD we multiply the vertex with a factor:
\begin{equation}
F_{N\Delta\gamma}(q^2) = 
\frac{m_{\rho}^2}{m_{\rho}^2 - i m_\rho \Gamma_\rho(q^2) - q^2}.
\end{equation}
In principle we would also need to include a pion-cloud contribution like
we did for the nucleon formfactor. 
However, we are not aware of such a calculation, and we have to ignore this
contribution for practical reasons. 

In Fig. \ref{results_1} we show for various calculations the resulting 
dilepton spectrum at a laboratory energy of 2 GeV,  
reachable e.g. by the COSY facility. 
The full line is obtained with no formfactors at all with only the 
nucleonic diagrams included in the calculation. 
The long-dashed line is found after including the pionic-cloud contributions
to the formfactor, for all practical purposes it coincides with the 
result without formfactors. 
There is surprisingly little effect of these contributions;
off-shell effects of the nucleon formfactor appear to be of minor
importance as far as dilepton production is concerned.
This can also be used as an ad-hoc justification for ignoring the
pion-cloud contributions on the $N\Delta\gamma$-vertex.

Only after including the VMD contribution do we find a significantly 
changed result, represented by the dotted line. 
This is of course not surprising, roughly speaking the VMD 
contribution 'multiplies' the
vertex with a $q^2$-dependent factor, which has its maximum somewhere 
in the region $M \sim m_\rho,m_\omega$. 
The broad width of the $\rho$ gives rise to an enhancement of the 
cross-section over a broad region of invariant masses. 
On top op this we observe a sharp peak due to the $\omega$.

We explored the effect of the phases between the 
VMD terms by using the phase as found for the pion formfactor 
\cite{om_phase}, $\phi = 116^o$. 
At distances of $\sim$ 20 MeV from the $\omega$-peak we 
find interference effects up to 25 \%. 
The height of the peak is not affected, there 
the $\omega$ totally dominates the spectrum. 
Far away from the peak the interference effects diminish, at $q^2 = 500$ MeV
we find only 5 \%.

We also performed a calculation which included the diagrams with $\Delta$ 
intermediate states.
Without any formfactors we obtain the short-dashed line, especially at lower 
invariant masses we see that the $\Delta$ contributions dominate the
cross-section. 
However, at higher invariant masses both the $\Delta$ and nucleonic
contributions are of equal importance. 
Including the formfactors on all vertices as describe above, we find the 
dash-dotted curve. 
We again observe the broad enhancement due to the $\rho$-meson.
The sharp $\omega$-peak is still clearly visible, although it exclusively
arises from the nucleonic contributions. 
This sharp peak provides an excellent probe to experimentally verify
the validity of VMD in the timelike region for the nucleon, provided 
the experimental resolution in invariant mass is good enough. 
The fact that the $\omega$ peak is exclusively due to the nucleon
formfactor makes it possible to separate such a measured peak from
the $\Delta$ contributions and assign it exclusively
to the nucleonic contributions.

Finally, the dash-dot-dot line is the result found when using a 
pseudo-vector coupling instead of a pseudo-scalar coupling for the 
pion in the effective OBE interaction.
No formfactors were included, the curve should be compared with the
full curve, which was obtained with a pseudo-scalar coupling for 
the pion. 
This effect is almost of the same order as that of the VMD formfactor. 
The difference arises from the magnetic part of the
$pp\gamma$ vertex: one can easily show that the contribution to the
cross-section of the electric part is independent of this choice
\cite{Schaefer_2}. 
Recall that no contact graphs due to the pseudo-vector coupling 
appear on isospin conserving vertices as we have in our model.
This also implies that our conclusion for the $F^{\pm+}_{1,3}$ 
formfactors are not affected by either taking the pseudo-vector or
pseudo-scalar coupling for the pion.
 
In conclusion, we investigated the effects of the electric formfactor
on dilepton production off $pp$ collisions. 
We used a model of the formfactor that describes it in terms of 
a nucleon surrounded by a pion cloud plus vector meson dominance and contains
an explicit off-shell dependence. 
We found that the off-shell dependent pion-cloud contribution hardly 
affects the result,
while the VMD part of the formfactor does increase the cross-section
significantly.
Assuming VMD leads to a sharp $\omega$-peak, which should be experimentally
observable, provided one has sufficient resolution for the invariant mass.

This work was supported by GSI Darmstadt, KFA Juelich and BMBF.

\begin{figure}
\caption{The various diagrams included in the current, the intermediate
state can either be a nucleon or a $\Delta$-isobar.}
\label{diagrams}
\end{figure}

\begin{figure}
\caption{
Differential cross-section for dilepton pair production off pp-collisions
at 2 GeV laboratory energy as a function of the invariant mass $M$. 
The full line is the result with no formfactors, the long-dashed line the
one with only the formfactor arising from the pion cloud included.
The dotted line is obtained including the VMD contribution on top of
the pion cloud contribution. 
Including the $\Delta$-contributions we find without formfactors the 
short-dashed line, incorporating all formfactors we obtain the dash-dotted
line. 
The dash-dot-dot line is found when using a pseudo-vector coupling for
the pion in stead of a pseudo-scalar coupling without including formfactors.
}
\label{results_1}
\end{figure}

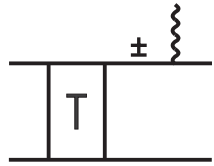
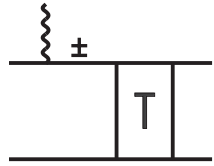
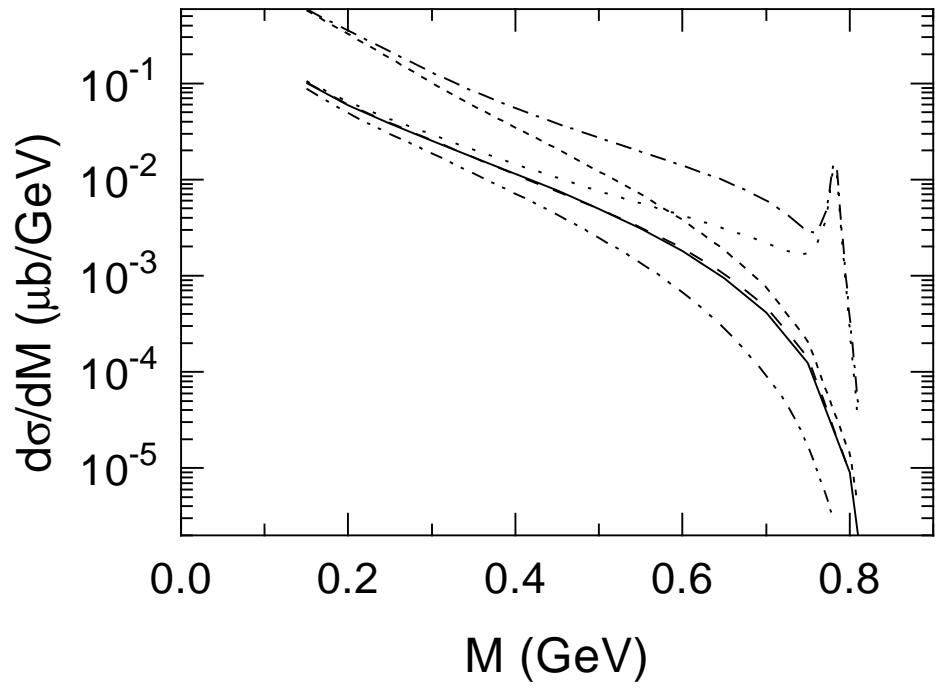
\end{document}